\documentclass{jfm}
   \usepackage{graphicx}
   \usepackage{amsmath,amssymb}
   \usepackage{natbib}
   \usepackage{latexsym}
   \usepackage{url}
   \usepackage{bm}
   \usepackage{color}
   \usepackage{epstopdf}
   \usepackage[margin=0cm,justification=justified]{caption}

   \graphicspath{{../figures}}

\title{Scaling and mechanism of the propagation speed of {\color{black}the upstream} turbulent front in pipe flow}
\author{Haoyang Wu\aff{1}, 
 \and Baofang Song\aff{2} \corresp{\email{baofang.song@pku.edu.cn}}}

\affiliation{\aff{1}Center for Applied Mathematics, Tianjin University, Tianjin 300072, China
            \aff{2} State Key Laboratory of Turbulence and Complex system, College of Engineering, Peking University, Beijing 100871, China
           }

\date{\today}
\begin{document}
\maketitle
\begin{abstract} 
Scaling and mechanism of the propagation speed of turbulent fronts in pipe flow with the Reynolds number has been a long-standing problem in the past decades.
Here, we derive an explicit scaling law of the upstream front speed, which approaches to a power-law scaling at high Reynolds numbers, and explain the underlying mechanism. Our data show that the average wall distance of low-speed streaks at the tip of the upstream front, where transition occurs, appears to be constant in local wall units in the wide bulk-Reynolds-number range investigated, {\color{black}between 5000 and 60000}. By further assuming that the axial propagation of velocity fluctuations at the front tip, resulting from streak instabilities, is dominated by the advection of the local mean flow, the front speed can be derived as an explicit function of the Reynolds number. The derived formula agrees well with the measured speed by front tracking. {\color{black}Our finding reveals the relationship between the structure and speed of a front, which} enables to obtain a close approximation of the front speed based on a single velocity field without having to track the front over time.
\end{abstract}

\indent 
\section{Introduction}
\begin{figure}
\centering	
 \includegraphics[width=0.5\linewidth]{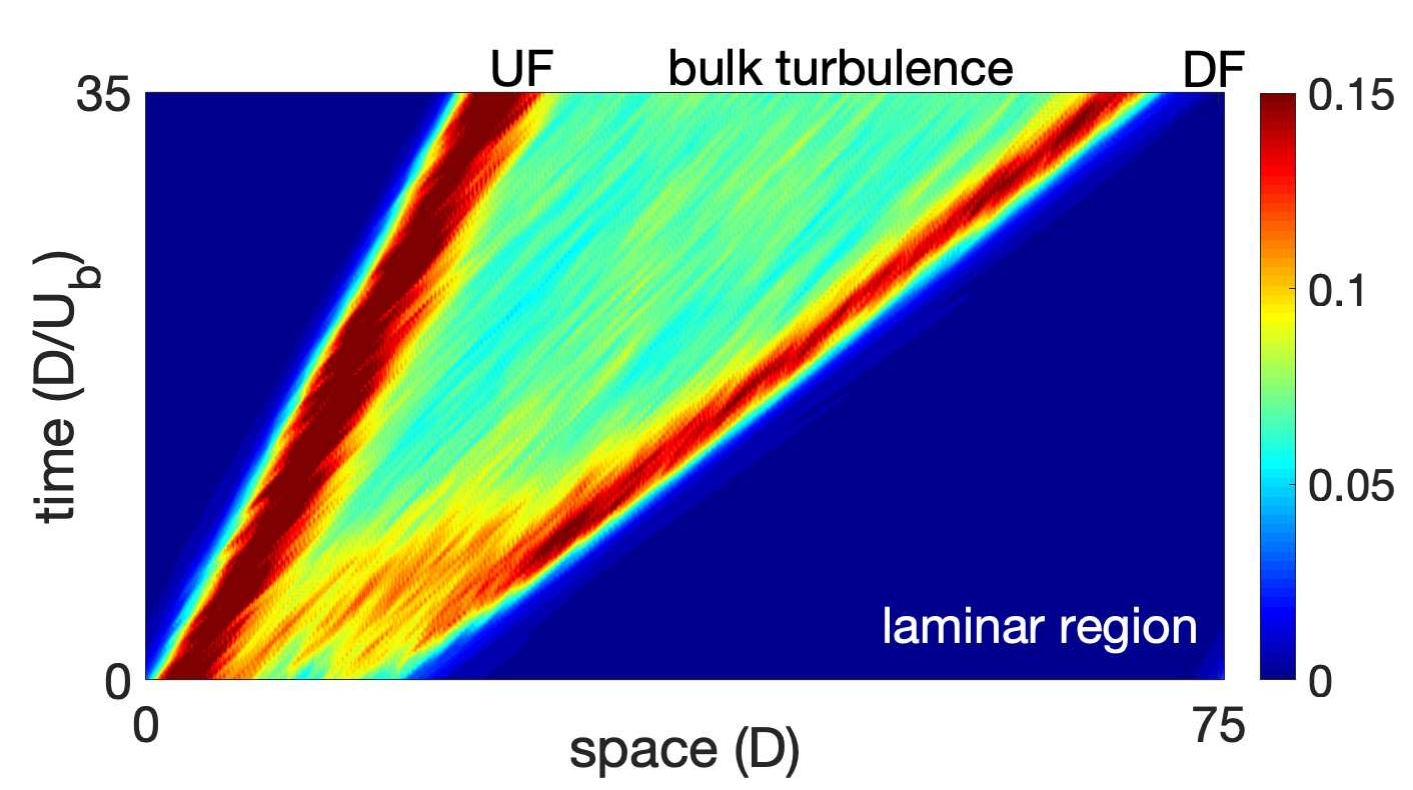}
\caption{\label{fig:spacetime_plot} The expansion of a turbulence at $Re=5000$. Contours of the magnitude of  transverse velocity  (averaged over the pipe cross-section) are plotted in the space (pipe axis) and time plane. {\color{black}The length unit in space is pipe diameter $D$ and time unit is $D/U_b$ with $U_b$ being the bulk speed of the flow.} The main stream is from left to right, and time is vertically up. Red color shows highly turbulent and blue color shows laminar regions. Different slopes of the two red stripes (fronts) indicate different front speeds, i.e. the turbulent region expands.}
\end{figure}

Front formation and propagation are important processes in nonlinear systems involving reaction, diffusion and advection, such as combustion, neural systems, epidemics and turbulent flows.
In pipe flow, a localized turbulent region expands via the propagation of the upstream front (UF) and downstream front (DF) into the laminar region, where the flow is stable to infinitesimal perturbations \citep{Meseguer2003b, ChenWeiZhang2022}. See FIG. \ref{fig:spacetime_plot} for an illustration. The front speeds determine the expansion rate of the turbulent region and consequently the wall friction. Therefore, front speed is an important characteristic of pipe flow turbulence and, together with the front structure, has been the subject of many studies in the past six decades \citep{Lindgren1957, Lindgren1969, Wygnanski1973, Darbyshire1995, Shan1999, Durst2006, Nishi2008, vanDoorne2008, Duguet2010b, Holzner2013, Barkley2015, Barkley2016, Song2017, Schlatter2019, Wang2022, Chen2022}. Yet the mechanism that determines the front speed and the scaling of the speed with the Reynolds number (Re) remains largely unclear to date. 

Most relevant studies focused on the narrow regime of transition from localized puffs to expanding turbulent states, i.e. slugs, at relatively low Reynolds numbers of O($10^3$). Due to difficulties of measuring front speed at high $Re$ \citep{Chen2022}, especially for the DF, only \citet{Wygnanski1973} and \citet{Chen2022} considered higher $Re$ at O($10^4$). 
Reasonable agreement has been obtained among existing measurements for the UF speed, showing a monotonically decreasing trend as $Re$ increases (see \citet{Chen2022} and \citet{Avila2023} for a latest literature review). However, \citet{Wygnanski1973} and \citet{Chen2022} reported opposite speed trends above $Re \simeq 10000$ for the DF. By direct numerical simulations (DNS) up to $Re=10^5$, \citet{Chen2022} reported fits $\tilde c_{UF}=0.024+(Re/1936)^{-0.528}$ for the UF and $\tilde c_{DF}=1.971-(Re/1925)^{-0.825}$ for the DF, and argued for the monotonic trends of the two front speeds with $Re$ at high Reynolds numbers. Although fitting the measured speeds well, these are pure data fits with a prescribed form but the underlying mechanism was left unexplained. In these fits and throughout this paper, the reference length and velocity are pipe diameter $D$ and bulk speed $U_b$, respectively. The Reynolds number is defined as $Re=U_bD/\nu$ where $\nu$ is the kinematic viscosity. We will only consider the UF speed in this work.

Other than direct measurements of the front speed by tracking the front along the pipe axis, as in most studies, a few theoretical attempts have been made. For example, based on an energy flux analysis of the front region, without considering the dynamic transition process, \citet{Lindgren1969} predicted an asymptotic speed of 0.69 as $Re\to \infty$. However, this prediction was questioned by both experimental \citep{Wygnanski1973} and numerical \citep{Chen2022} measurements, which showed much lower speeds than the asymptotic prediction of \citet{Lindgren1969}. 
\citet{Barkley2015} and \citet{Barkley2016} used a theoretical model to investigate the front in the transitional regime, which captures the large-scale dynamics of the fronts successfully.
The asymptotic analysis explains the front speed as a combination of the advection of the bulk turbulence and a propagation with respect to the bulk turbulence (see FIG. \ref{fig:spacetime_plot}).
However, as a generic model for one-dimensional reaction-diffusion-advection systems, the model does not account for the three dimensionality of the transition for pipe flow and does not give an explicit relationship between the front speed and $Re$. Besides,
the $Re$ range considered was too narrow for a scaling with $Re$ to be established. Similar problems apply to the stochastic prey-predator model recently proposed by \citet{Wang2022} which nevertheless can reproduce the basic phenomenology of the front of pipe flow in the transitional regime. In a word, there is still a big gap between experiments and theory.

In this paper, our goal is to derive a scaling law of the front speed by accounting for the transition at the front.

\section{Results}
Our starting point is the observation of our earlier work \citep{Chen2022} that transition to turbulence continuously occurs at the tip of a front (see supplementary movie), maintaining a characteristic propagation speed and a characteristic shape of the front against the distortion of the mean shear. {\color{black}For an upstream front, the front tip refers to its upstream-most point.} We propose that an evaluation of the radial position of the transition point at the front tip is crucial for determining the front speed. The questions are how to quantitatively determine this position and relate it to the front speed. 
\begin{figure}
\centering	
 \includegraphics[width=0.95\linewidth]{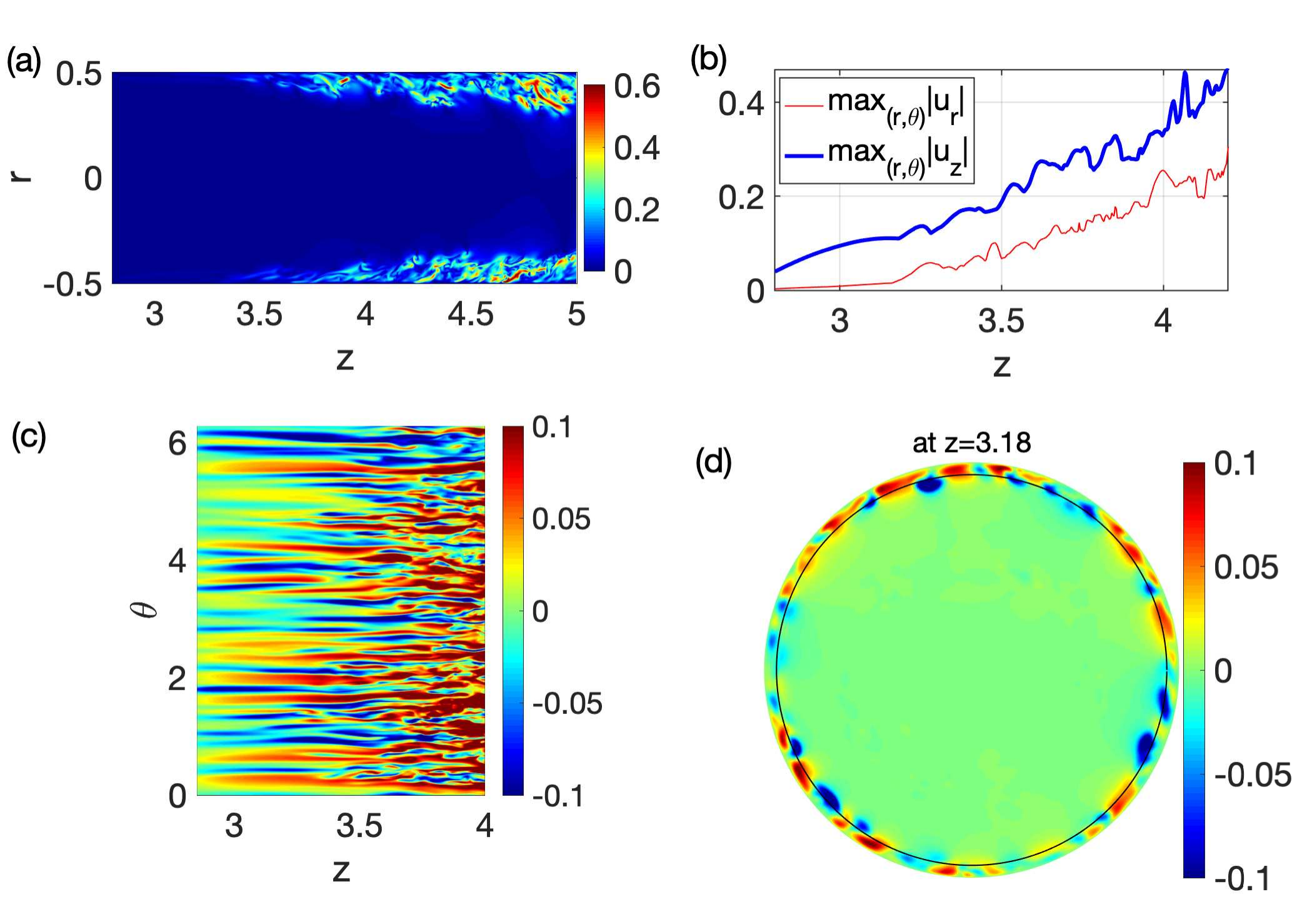}
\caption{\label{fig:streaks_at_UF} Structure of a UF. (a) The tip part (the upstream-most part) of a UF in a $z-r$ plane at $Re=40000$. Contours of the transverse velocity $\sqrt{u_r^2+u_{\theta}^2}$ are plotted in the $z-r$ crosssection. The main flow is from left to right, and blue region is low-velocity-fluctuation region and red region is high-velocity-fluctuation region. (b) The maximum of $|u_r|$ and $|u_z|$ in the $r-\theta$ crosssection plotted along the pipe axis. (c) Contours of $u_z$ in the $z-\theta$ plane at $r=0.47$, which show low-speed (blue) and high-speed (red) streaks. (d) Contours of $u_z$ in the $r-\theta$ crosssection at $z=3.18$. The circle, at $r=0.47$, shows the radial position of the $z-\theta$ plane in panel (c). This circle approximately shows the average position of low-speed streaks (blue spots) at this axial position.}
\end{figure}

Before presenting our results, the setup of the flow system and some notations should be explained. The flow is incompressible and constant-mass-flux driven, and is solved in cylindrical coordinates, where $(r,\theta,z)$ denote radial, azimuthal and axial coordinates, respectively, and $u_r$, $u_\theta$, and $u_z$ denote the respective fluctuating velocity components in the three directions. 
The axial length of the pipe domain is $17.5D$ for $Re\le 10000$ and $5D$ for higher $Re$.
The readers are referred to \citet{Chen2022} for more details about the simulation of a front in a short periodic pipe. 
 
We first show the flow structure at the tip of the UF. In figure \ref{fig:streaks_at_UF}(a), contours of the magnitude of transverse velocity fluctuation in the $z-r$ plane show that turbulence is concentrated near the wall and gradually spreads out toward the pipe center while going downstream (see the online movie for more details). 
Figure \ref{fig:streaks_at_UF}(b) shows the distribution of maximum $|u_z|$ and $|u_r|$ in the $r-\theta$ crosssection along the pipe axis. These curves also reflect that the flow is nonturbulent on the upstream {\color{black} side of the front tip} and turbulent on the downstream side. Figure \ref{fig:streaks_at_UF}(c) shows that the flow features nearly straight and streamwise-elongated low-speed (blue) and high-speed (red) streaks on the upstream. The flow structure is less regular on the downstream, indicating turbulence. {\color{black} These plots (especially figure \ref{fig:streaks_at_UF}b) suggest that} the tip of the UF should sit in the interval $z\in (3, 3.5)$ roughly. Figure \ref{fig:streaks_at_UF}(d) shows the contours of $u_z$ in the $r-\theta$ plane at $z=3.18$. Alternating high-speed (red spots) and low-speed streaks (blue spots) can be seen close to the wall, while the flow is laminar in the core region of the pipe.

Although there still lacks a quantitative description of the transition mechanism at the UF tip, the consensus seems to be that the transition is caused by instabilities of the low-speed streaks \citep{Shimizu2009, Duguet2010b, Hof2010} (which may consist of more fundamental substructures according to \cite{Jiang2020, Jiang2020b}). 
The instabilities here possibly coincide with those (either modal or nonmodal) proposed for explaining either subcritical transition or the self-sustaining mechanism of shear flow turbulence  \citep{Swearingen1987, Hamilton1995, Zikanov1996, Waleffe1997, Schoppa1998, Schoppa2002, Meseguer2003}.  
In the following, we will establish a connection between the low-speed streaks and the front speed based on a few hypotheses, the first of which reads

\begin{enumerate}
\item[H1]{\it The wall distance of transition point at the front tip, in local wall unit, is independent of the Reynolds number statistically. }\label{hyp:H1}
\end{enumerate}

This should be reasonable because the transition takes place near the wall so that the wall distance of the transition point can be expected to scale with the wall length, which is the only length scale that can be derived from viscosity and wall shear. We will verify this hypothesis by measuring the wall distance of low-speed streaks as a proxy of that of the transition point.

\begin{figure}
\centering	
 \includegraphics[width=0.99\linewidth]{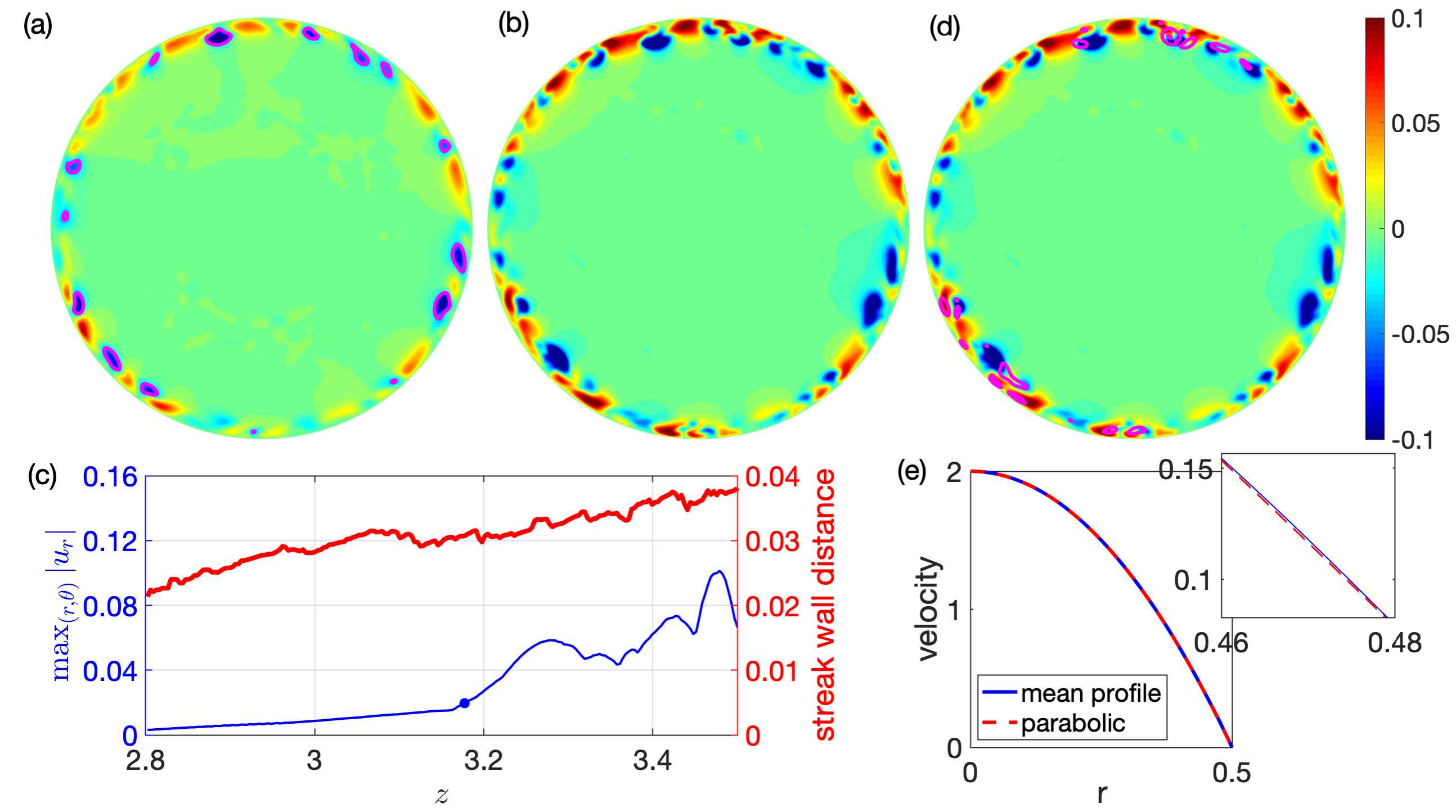}
\caption{\label{fig:streaks_profiles} Axial variation of the low-speed streaks (a-d) and mean velocity profiles near the front tip (e). The flow field is the same as that shown in FIG. \ref{fig:streaks_at_UF}. (a) Contours of $u_z$ at $z=3.0$. The contour level of $-0.04$ is plotted in magenta color to highlight the low-speed streaks. (b) Contours at $z=3.3$. 
(c) The variation of the average wall distance of streaks (bold red) and $\max_{(r,\theta)}|u_r|$ (thin blue) along the pipe axis.
(d) Also at $z=3.3$, and the contour level $0.025$ of the transverse velocity $\sqrt{u_r^2+u_\theta^2}$ is plotted in magenta color to highlight the transverse velocity fluctuations. (e) The mean velocity profile at $z=3.3$, i.e. $U(r)=<u_x(r,\theta, 3.3)>_{\theta}$, where $<\cdot>_\theta$ means average in the azimuthal direction. The parabolic profile is plotted as a broken line for comparison. {\color{black}The small window highlights the deviation between the two in the near wall region}.}
\end{figure}

At a turbulent front, the flow is axially developing so that low speed streaks are not parallel to the pipe wall but oblique, i.e. the wall distance varies along a streak. Figure \ref{fig:streaks_profiles}(a,b) suggest that low-speed streaks are gradually lifted up away from the wall while going downstream. 
To more quantitatively show this variation, we take the following approach to determine the wall distance of low-speed streaks at a given axial location. In an $r-\theta$ cross-section, low-speed streaks can be detected by setting a proper threshold in $u_z$, and regions enclosed by contour lines of the specified threshold can be considered as (the cross-sections of) low-speed streaks, see the magenta contour lines in Fig. \ref{fig:streaks_profiles}(a) with a threshold $-0.04$. See Appendix \ref{sec:threshold_for_streaks} for a discussion on the threshold selection. Then, the nominal wall distance of a streak can be defined as the wall distance of the minimum of $u_z$ within the streak.  
The average wall distance is calculated as the arithmetic mean of the wall distances of all the streaks detected in this pipe cross-section.
Fig. \ref{fig:streaks_profiles}(c) more quantitatively shows that the wall distance of streaks increases as going downstream.
Therefore, it is necessary to determine the axial position of the front tip for finally determining the wall distance of the streaks at the front tip. We use $\max_{(r,\theta)}|u_r|$, which is a function of $z$, as an indicator of the local flow state. This curve is smooth and slowly varying in the laminar region and wiggles around in turbulent region, see Fig. \ref{fig:streaks_at_UF}(b) and Fig. \ref{fig:streaks_profiles}(c). The axial location of the front tip can be estimated by the position separating the smooth and wiggling parts of the curve of $\max_{r,\theta}|u_r|$. We use an algorithm that detects abrupt changes of a curve for this purpose, which is built-in as the function {\it findchangepts} in MATLAB {\color{black}(see Appendix \ref{sec:abruptchangepts} for a brief description of the algorithm)}. The blue dot in Fig. \ref{fig:streaks_profiles}(c) shows the separating point determined using this algorithm.

\begin{figure}
\centering	
 \includegraphics[width=0.99\linewidth]{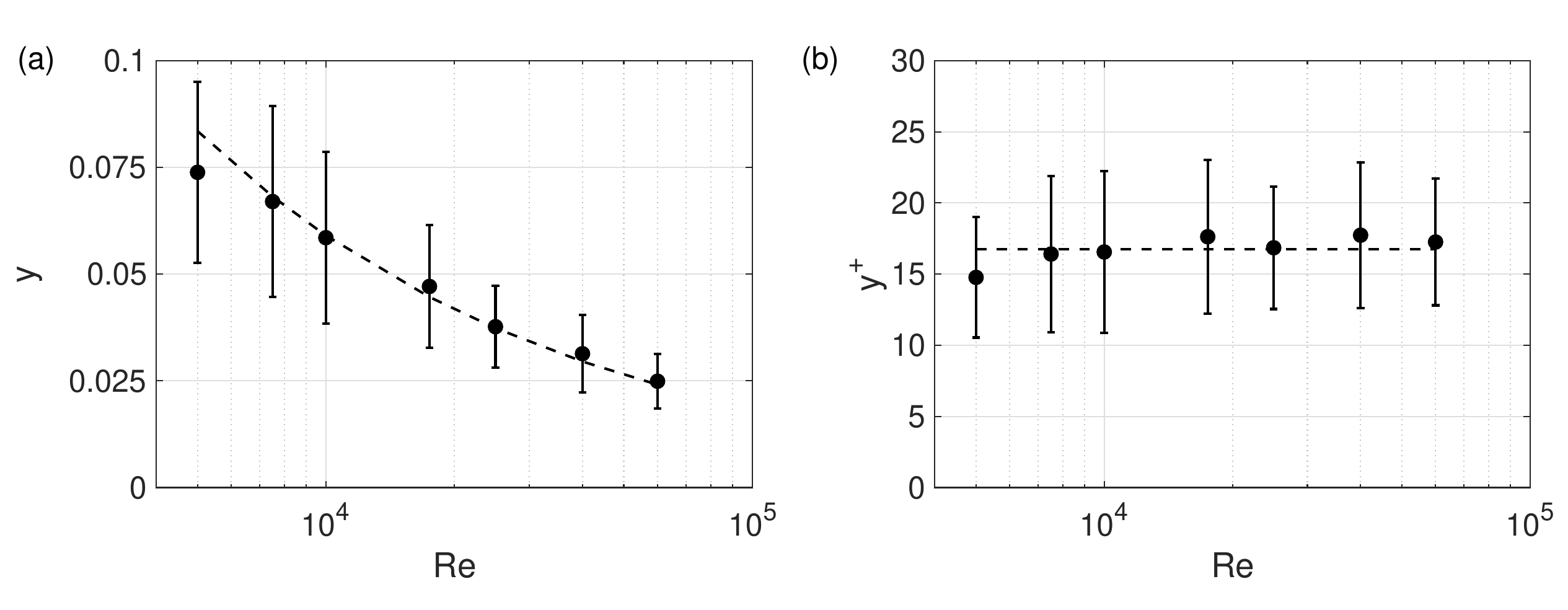}
\caption{\label{fig:streaks_position} The average wall-distance of the low-speed streaks at the tip of the UF. (a) The distance in the outer length unit. (b) The distance in the local wall unit. At each $Re$, about 10 to 20 velocity snapshots are collected, giving 100 to 200 low-speed streaks approximately for the statistics. The standard deviation is plotted as the errorbar. {\color{black} The dashed lines are equation (\ref{equ:wall_distance}) by taking $A=16.7$, which is the average of $y^+$ over all $Re$ shown in panel (b)}.
}
\end{figure}

Figure \ref{fig:streaks_position}(a) shows the average wall distance of streaks $y=0.5-r$ at the front tip in outer units. The larger the $Re$, the smaller the $y$, which can be expected. 
Figure \ref{fig:streaks_position}(b) shows $y^+$, the wall distance in local wall length unit $\sqrt{\nu/\tau_w}$, where $\tau_w$ is the local wall shear stress. Considering that the {\color{black}azimuthally averaged} velocity profiles at these axial locations are nearly parabolic (see FIG. \ref{fig:streaks_profiles}e), $\tau_w$ is simply approximated by the value of the parabolic profile. It appears that $y^+$ stays nearly constant in the wide $Re$ range considered, which supports our hypothesis H1 given the crucial role that  low-speed streaks play in the transition.

Assuming this $Re$-independence, we derive the scaling law of the speed of the UF as following. Taking the wall distance of the transition point at the front tip to be $y_F^+=A$, where $A$ is independent of $Re$. Then, in outer units, we have 
\begin{equation}\label{equ:wall_distance}
y_F=y_F^+/Re_{\tau}=A/Re_{\tau}.
\end{equation}
The local mean flow speed, i.e. {\color{black}the azimuthally-averaged streamwise velocity} at the {\color{black}radial position of the} transition point, in outer units, can be approximated by
\begin{equation}
U(y_F)\approx 2-8\left(0.5-y_F\right)^2=8\left(A/Re_{\tau}-A^2/Re_{\tau}^2\right),
\end{equation}
given that the mean velocity profile is nearly parabolic at the front tip.
As the relationship between $Re$ and $Re_\tau$ is
\begin{equation}
Re_{\tau}=\sqrt{-\frac{\mathrm{d}U(r)}{\mathrm{d}r}\vert_{r=0.5}Re}~~= 2\sqrt{2Re}
\end{equation}
for a parabolic velocity profile, we have
\begin{equation}
U(y_F) \approx 2\sqrt{2}ARe^{-0.5} - A^2 Re^{-1}.
\end{equation}
Now it comes to our further hypotheses:
\begin{enumerate}
\item[H2]{\it The wall distance of velocity perturbations at the front tip, resulting from streak instabilities, can be closely approximated by the  wall distance of the streaks.} \label{hyp:H1}
\item[H3] {\it {\color{black}The front speed is determined by the axial propagation speed of these velocity perturbations, which approximately equals the local mean flow speed.}}\label{hyp:H2}
\end{enumerate}
H2 should be reasonable, especially when  perturbations appear at the flanks of the streaks. In fact, the data seems to support this hypothesis, see FIG. \ref{fig:streaks_profiles}(d) where most of the strong-perturbation region, enclosed by magenta contour lines, seems to be at the flanks of the low-speed streaks. 
{\color{black} H3 is based on our presumption that streak instability generates streamwise vortices, which further generate streaks while being advected downstream, seeding new transition and closing the self-sustaining cycle of the dynamics at the front tip.} The propagation of vortical structures, at least in fully developed wall turbulence above the viscous sublayer, was shown to be dominated by the advection of the local mean flow \citep{Alamo2009, Pei2012, Wu2008}. H3 should be reasonable if our presumption is reasonable, {\color{black}but the detailed dynamics at the front tip certainly needs further studies and is out of the scope of this paper}.

Following these hypotheses, we finally have an approximation of the front speed as
\begin{equation}\label{equ:approximate_cUF}
c_{UF}\approx U(y_F) \approx 2\sqrt{2}ARe^{-0.5} - A^2 Re^{ -1},
\end{equation}
and an asymptotic approximation at large $Re$
\begin{equation}\label{equ:asymptotic_cUF}
c_{UF} \approx U(y_F)\approx 2\sqrt{2}ARe^{-0.5},
\end{equation}
{\color{black} where $A$ can be approximated by the wall distance of low-speed streaks at the front tip.}

\begin{figure}
\centering	
 \includegraphics[width=0.99\linewidth]{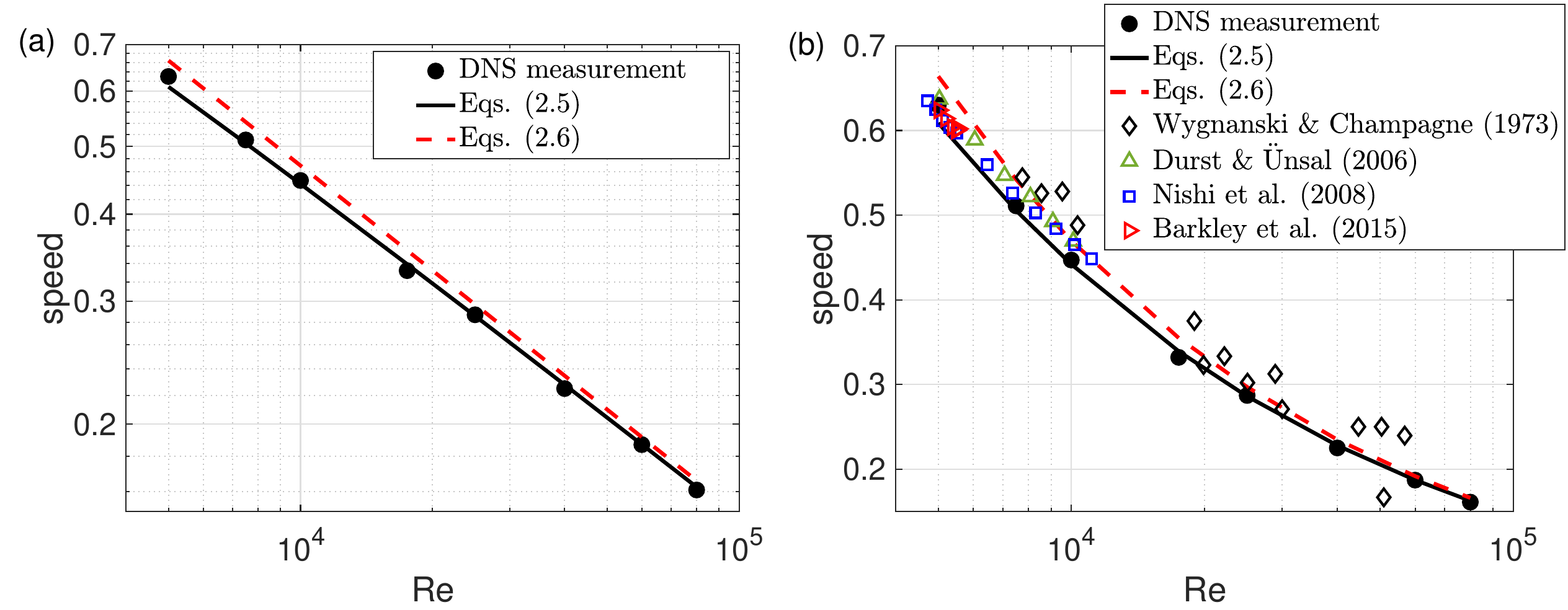}
\caption{\label{fig:front_speed} Comparison of derived and measured front speeds. The circles show the speeds measured by front tracking using DNS \citep{Chen2022}. 
The solid black line shows the approximation Eqs. (\ref{equ:approximate_cUF}) with $A=16.7$, which is the average of the $y^+$ (black circles in FIG. \ref{fig:streaks_position}b) over all $Re$'s. The dashed red line shows the approximation Eqs. (\ref{equ:asymptotic_cUF}) with $A=16.7$ also. Panel (b) is the same plot in linear scale for the speed {\color{black} where some data sets from the literature falling in this $Re$ range are also included.}}
\end{figure}

Figure \ref{fig:front_speed} concludes the speed measurements and our derivation. The filled circles are the DNS data from \cite{Chen2022} (up to $Re=60000$) and the open symbols show the literature data in the $Re$ range investigated here. In order to show that the formula is predictive, DNS at $Re=80000$ is performed here, and the front speed is measured by front tracking and plotted as a filled circle also. 
The black solid line shows our derivation (\ref{equ:approximate_cUF}) by setting $A=16.7$, which is the average of $y^+$ of streaks at all $Re$'s as shown in figure \ref{fig:streaks_position}b. The relative error of the prediction is on the level of a few percent compared to the DNS measurement.
The red dashed line shows the asymptotic speed (\ref{equ:asymptotic_cUF}) with the same $A$. {\color{black}Some former experimental measurements are also included in the figure.}
It should be noted that this formula can also be considered as a model for the front speed with only one parameter $A$, which has a physical meaning and, more precisely, should be interpreted as the wall distance of the transition point at the front tip.
This formula can be used for other $Re$'s after calibrating the parameter $A$ at one $Re$ with the measured front speed.

Now we revisit the fit $\tilde c_{UF}=0.024+(Re/1936)^{-0.528}$ by \citet{Chen2022}.
This was obtained by assuming a form of $a+bRe^\beta$ without an explanation of the underlying physics. In other words, this form is not unique. Besides, the small constant 0.024 implies that the front speed would not approach zero as $Re$ approaches infinity, which was unexplained and seems counter-intuitive. It is probably just a result of measurement errors and the specific prescribed form of the fit.
In contrast, our derivation (\ref{equ:approximate_cUF}) makes no assumption on the specific form of the formula. It follows naturally from the dynamics we observed at the front tip with a few hypothetical but reasonable assumptions of the physics.

Our derivation (\ref{equ:approximate_cUF}) may suffer larger errors at lower $Re$. The low-speed streaks would be larger in transverse size at lower $Re$, therefore, the position of a streak estimated simply by the position of the minimum of $u_z$ in each streak becomes less representative. Besides, the streak position may not exactly coincide with the position of velocity fluctuations resulting from the streak instability. But these positions are close to each other at sufficiently high $Re$ so that our derivation will be more accurate.  


As for the DF, the front speed is probably determined by the advection of the local mean flow at the front tip also. However, transition to turbulence occurs close to the pipe center \citep{Chen2022} and the transition may not be triggered by streak instabilities as known for near-wall turbulence. Therefore, the location of the transition point may not scale with the wall length unit, and
cannot be explicitly related to $Re$ as shown here for the UF at the present. This problem has to be left for future studies. 

\section{Conclusions}
In summary, the speed of the UF of pipe flow turbulence was derived as an explicit function of $Re$ based on the dynamics at the front tip. To our knowledge, this is the first of such since the seminal measurements and theoretical analysis of \cite{Lindgren1957, Lindgren1969} about six decades ago.
The agreement with speed measurements (see FIG. \ref{fig:front_speed}) suggests that the mechanism proposed here captures the core of the physics, i.e. the front speed is largely determined by the advection of velocity fluctuations by the local mean flow at the front tip where transition takes place.
This mechanism may also apply to turbulent fronts in other shear flows where turbulence propagates into subcritical laminar flow region. Although the local mean flow is {\color{black}different in} higher dimensions such as planar shear flows (see, e.g. \citet{Duguet2013, Tao2018, Tuckerman2020, Klotz2021}), our work will be helpful for elucidating the physics of front propagation in those flows.

\section*{Acknowledgments}
The authors acknowledge the financial support from the National Natural Science Foundation of China under grant numbers 12272264 and 91852105. The work is also supported by ``The Fundamental Research Funds for the Central Universities, Peking University". Simulations were performed on Tianhe-2 at the National Supercomputer Centre in Guangzhou and Tianhe-1(A) at the National Supercomputer Centre in Tianjin.

\noindent Declaration of Interests. The authors report no conflict of interest.

\appendix
\section{Threshold for detecting streaks}
\begin{figure}
\centering	
 \includegraphics[width=0.99\linewidth]{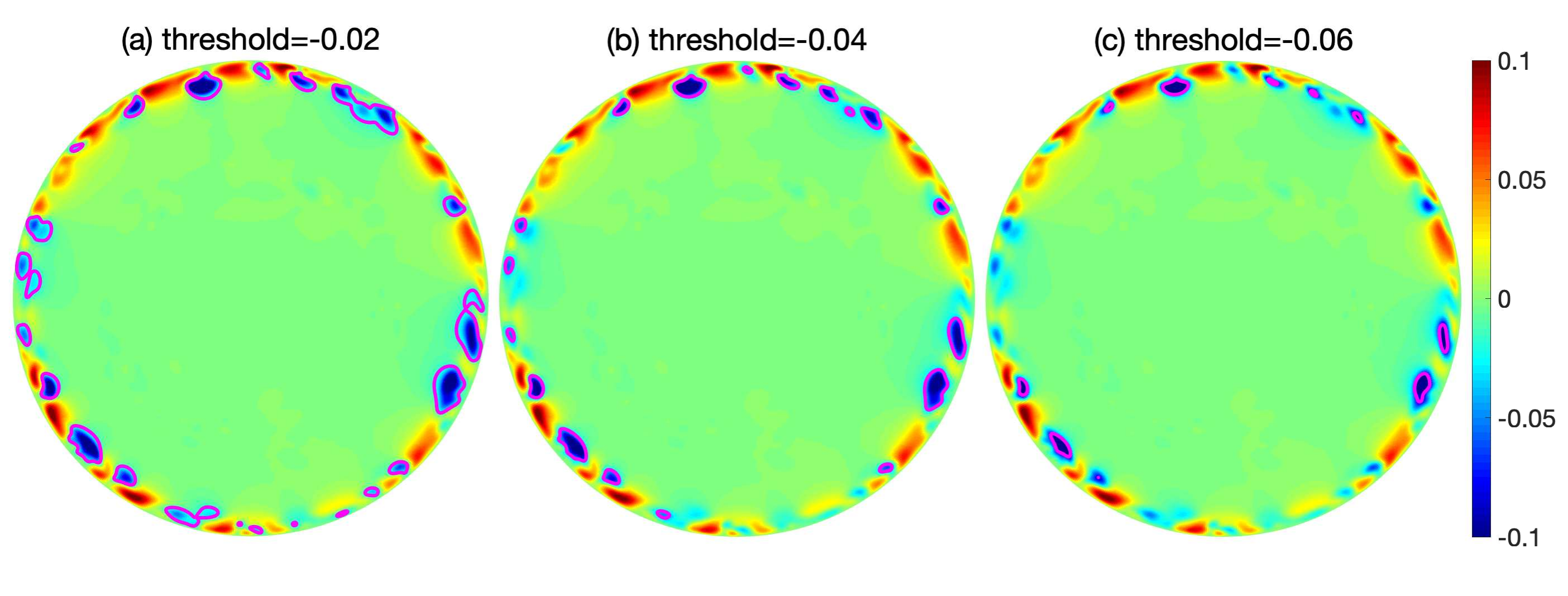}
  \includegraphics[width=0.5\linewidth]{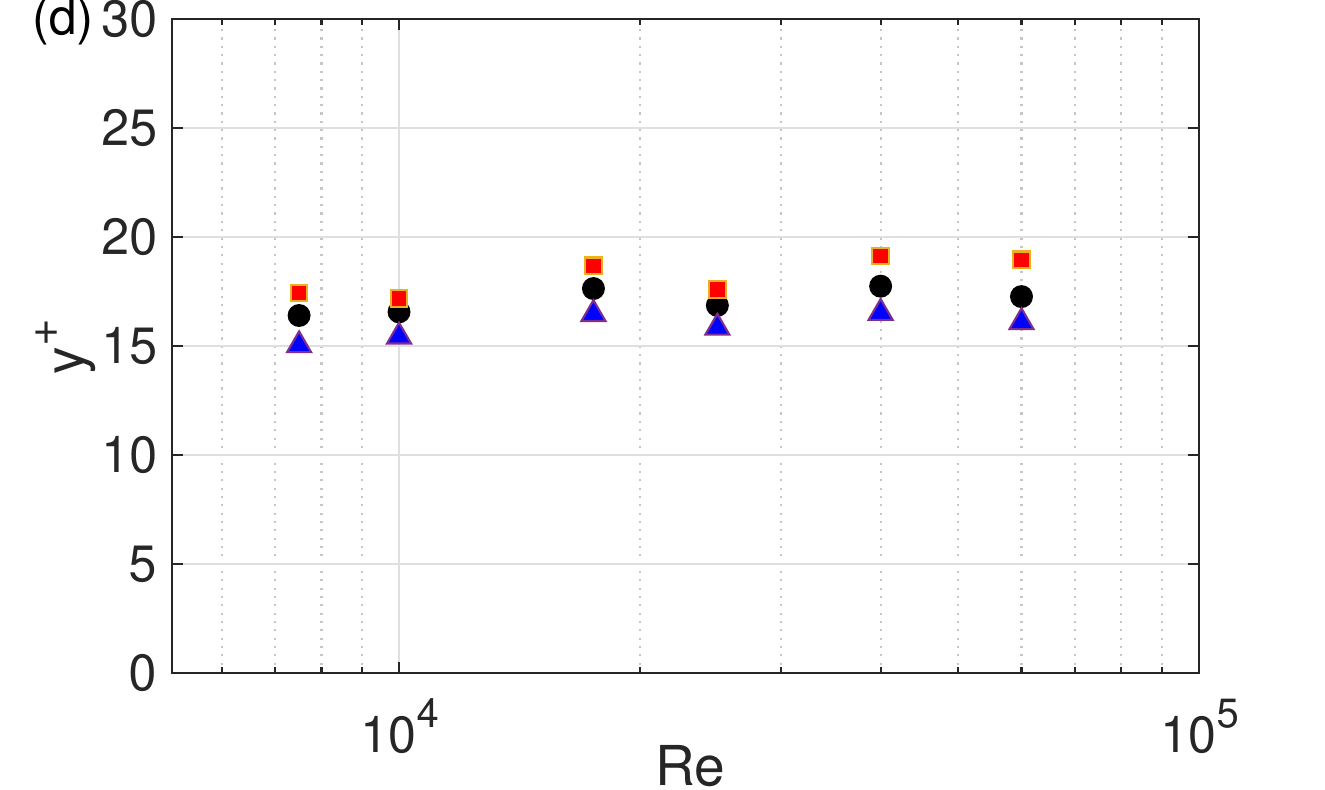}
\caption{\label{fig:threshold_streaks} Thresholds for detecting low-speed steaks. Contours of streamwise velocity fluctuation in the $r-\theta$ crosssection for $Re=40000$ as also shown in FIG. 2 of the main text. Contour levels of $-0.02$, $-0.04$ and $-0.06$ are plotted as magenta lines. (d) The average wall distance of low-speed streaks in wall units determined with thresholds of $-0.02$ (blue triangles), $-0.04$ (black circles) and $-0.06$ (red squares) in $u_z$ for detecting the streaks.
}
\end{figure}

{\label{sec:threshold_for_streaks}}
The results presented in the main text takes the threshold of $-0.04$. Here we explain the selection of this value. Figure  \ref{fig:threshold_streaks} shows the contour levels of $-0.02$, $-0.04$ and $-0.06$ in the $r-\theta$ crosssection at $z=3.18$ (the same position as shown in FIG 2(d) of the main text), plotted as magenta lines. It can be seen that $-0.02$ cannot very well separate adjacent streaks, whereas $-0.06$ may miss out many streaks. We checked multiple velocity snapshots and Reynolds numbers and found it is often the case. The threshold $-0.04$ is a reasonable choice because, in most cases, it separates streaks well and is able to detect most of low-speed streaks. 

It can be expected that the value of this threshold will affect the average position of the streaks. A higher threshold may drop out weaker streaks and only retain stronger streaks. Stronger streaks are often more lifted up away from the wall (can be seen in Figure \ref{fig:threshold_streaks}), and therefore, a higher threshold will give a larger average wall distance of the streaks. Here we measured the average wall distance $y^+$ of the streaks determined using thresholds $-0.02$ and $-0.06$, see the blue triangles and the red squares, respectively, in Figure \ref{fig:threshold_streaks}(d). 
It can be seen that $-0.02$ gives slightly lower and $-0.06$ gives slightly higher $y^+$ compared to the black circles (with a threshold of $-0.04$ for detecting streaks). 
But the important point is that the $y^+$ also appears to be a constant in the $Re$ range considered using either threshold for detecting the streaks. 

{\color{black}{
\section{The algorithm for determining the axial location of the front tip}{\label{sec:abruptchangepts}}
In the main text, we use the algorithm that is built in as the function {\it findchangepts} in MATLAB to detect abrupt changes in a signal sequence $[x_1, x_2, ..., x_n]$. The key is to minimize the following target function  
\begin{equation}
J(k)=\sum_{i=1}^{k-1}\left(x_i-\text{mean}([x_1, x_2, ..., x_{k-1}])\right)^2+\sum_{i=k}^n\left(x_i-\text{mean}([x_k,x_{k+1},...,x_n])\right)^2
\end{equation}
by modifying the index $k$. The resulted $k$ is regarded as the separation point of the slowly-varying and abruptly varying-parts of the sequence. The data sequence of $\max_{r,\theta}|u_r|$, containing both the laminar part and turbulent part on the upstream and downstream  sides of the front tip, respectively, is fed as the input. The output will be taken as the point separating the laminar part and turbulent part of the curve, which we define as the axial location of the front tip. The readers are referred to the documentation of MATLAB for more details about the algorithm. }}

\bibliographystyle{jfm}
\bibliography{references}

\end{document}